\documentclass[12pt,dvips]{article}

\usepackage{amsmath,amssymb,exscale}
\usepackage{array,multicol}
\usepackage{afterpage,float,flafter}
\usepackage{epsfig,rotating,pifont,fancybox}
\usepackage{cite}
\makeatletter
\@addtoreset{equation}{section}
\makeatother

\title{
\vspace*{-0.8cm}
\vspace{1cm} 
\Large\textbf{A note on CFT dual of RS model with gauge fields in bulk}
\vspace*{.5cm}
\author{\large 
\textbf{
K.~Agashe\footnote{email: kagashe@pha.jhu.edu}} 
\mbox{ } and 
\large \textbf{A.~Delgado\footnote{email: adelgado@pha.jhu.edu}
}\\
\emph{
Department of Physics and Astronomy} \\ 
\emph{Johns Hopkins University} \\ 
\emph{3400 North Charles St}. \\ 
\emph{Baltimore, MD 21218-2686}}}

\date{}
\begin{document}
\maketitle
\thispagestyle{empty}
\vspace*{.5cm}
  
\begin{abstract}

It has been conjectured that the (weakly coupled) Randall-Sundrum (RS)
model with 
gauge fields in the bulk is dual to
a (strongly coupled) $4D$ conformal
field theory (CFT) with an 
UV cut-off and in which
global symmetries of the CFT are gauged. We 
elucidate 
features
of this dual CFT which are 
crucial for a 
complete understanding of the proposed duality.
We 
argue that the limit of no (or small) 
brane-localized kinetic term for 
bulk
gauge field 
on the RS side
(often studied in the literature) is dual to
no
{\em bare} kinetic term for the gauge field which is coupled
to the CFT global current. In this limit, 
the 
kinetic term
for this gauge field in the dual CFT
is 
``induced'' by CFT loops.
Then, this 
CFT loop contribution to the gauge field 
$1$PI two-point function
is dual
(on the RS side) to the 
{\em full} 
gauge propagator 
(i.e., 
including 
the 
contribution 
of Kaluza-Klein {\em and} zero-modes) 
with both external points on the
Planck brane. 
We also emphasize
that {\em loop} corrections to the gauge coupling
on the RS side are dual to {\em sub-leading} effects
in 
a large-$N$ expansion on the CFT side; 
these sub-leading corrections to the gauge coupling in the dual CFT
are (in general) sensitive to the strong dynamics of the
CFT.

\end{abstract} 
  
\newpage 
\renewcommand{\thepage}{\arabic{page}} 
\setcounter{page}{1} 
  
\section{Introduction}

The Randall-Sundrum (RS) proposal of a warped extra dimension \cite{rs} is
interesting from both 
theoretical and phenomenological points of view, for example,
the RS1 model can solve the Planck-weak hierarchy problem. 
%
For the case of RS1, the
extra dimension is an orbifolded circle  
of radius $r_c$
with the 
Planck (or UV) brane at $\theta=0$ and the TeV (or IR) brane at 
$\theta = \pi$. The geometry is a compact slice of AdS
(with curvature scale, $k$ of order $M_4$, the $4D$ Planck mass):
\begin{eqnarray} 
ds^2 & = & e^{-2k r_c |\theta|} \eta_{\mu \nu} dx^{\mu} 
dx^{\nu} + r_c^2 d \theta ^2,  \;
- \pi 
\leq \theta \leq \pi
\nonumber \\ 
 & = & \frac{1}{(kz)^2} \left( \eta_{\mu \nu} dx^{\mu} 
dx^{\nu} + dz^2 \right), \; z_{UV}
\leq z \leq z_{IR}. 
\end{eqnarray} 
%
In terms of the coordinate
$z \equiv 1/k \; e^{k r_c |\theta|}$, 
the Planck brane is at 
$z_{UV} = 1/k$, the TeV brane is at $z_{IR} = 1/k \; 
e^{k \pi r_c}$ and, in order to solve the Planck-weak hierarchy 
problem, $r_c$ is chosen so that $1 / z_{IR} \sim $TeV. The
RS2 model corresponds to $r_c \rightarrow \infty$.

The AdS/CFT correspondence \cite{maldacena, gubser1, witten1, review}
suggests that RS models
are dual to deformed conformal field theories (CFT's) 
\cite{adscftrs2, nima, rattazzi, perez}.
In this paper, we are interested in the CFT dual of RS model
with gauge fields in the bulk \cite{hewett, pomarolplb, others}. 
In this case, the dual is
a CFT with a UV cut-off and with global symmetries 
of the CFT
gauged by a
gauge field {\em external} to the CFT \cite{witten1,nima}.
In this paper, we study
aspects 
of this dual CFT
which are 
required for a better
understanding of the AdS/CFT correspondence as applied to this RS model.
%
%
Although these CFT duals have been
discussed before, the 
properties we focus on have not been
studied in detail before.

The central point of our study
is that the {\em bare} kinetic term 
(or kinetic term at the UV cut-off) for gauge field external to
the CFT 
is to be identified
with Planck
brane-localized kinetic term for (bulk) gauge field on 
the RS side. As a check, we show that 
(the form of) the
propagator for gauge field coupled to the global current of the dual
CFT (including bare kinetic term {\em and} 
CFT loop correction) agrees with
the 
gauge propagator in RS model with both external
points on the Planck brane (including the effects of bulk {\em and} 
brane-localized kinetic term).

This implies that the limit of
no (or small) (Planck) brane-localized kinetic term on the RS side
is dual to no {\em bare} kinetic term for gauge field 
coupled to global current of the CFT,  
so that
the kinetic term for this gauge field is ``induced'' by CFT loops. 
This 
``induced gauge theory''
is similar to CFT loops inducing gravity (rather inverse Newton constant)
in the limit of very large (or infinite)
{\em bare} gravitational constant 
\cite{nima, rattazzi, hawking} -- 
we 
extend this
analysis 
to the case of the gauge field in order to have a 
complete
understanding of the AdS/CFT duality in this case.  
We also study
in more detail
the implications
of induced gauge theory and induced gravity. 
For example, 
in the induced gauge theory (or gravity) limit, the
CFT contributions to the external gauge field (or graviton) 
$1$PI two-point function
are dual (on the RS1 side)
to the contribution of zero {\em and} Kaluza-Klein (KK) modes to the Planck 
brane-to-Planck brane propagator.

Another consequence 
of 
induced kinetic term for gauge field external
to the dual CFT
is large
kinetic mixing between this gauge field 
and CFT bound states 
\cite{nima}. This 
leads to a dual interpretation of 
{\em flat} profile of gauge field zero-mode and {\em sizable}
coupling of gauge KK modes
to Planck brane fields. We 
compare the case of gauge field
to that of gravity, where we argue that
this mixing (between graviton and CFT fields) is small even in the
induced gravity limit. This matches with
{\em localization} of graviton
zero-mode near Planck brane and {\em weak} coupling of graviton
KK modes to Planck brane fields.
We also give a 
dual interpretation of coupling of gauge KK modes to {\em TeV} brane fields.

The AdS/CFT correspondence 
says that
tree-level
effects
in AdS are dual to leading effects in a large-$N$ expansion on the
CFT side (for example, if the
CFT is a $SU(N)$ gauge theory), whereas
{\em loop} effects on AdS side are
dual to {\em sub-leading} large-$N$ effects on the CFT side
\cite{maldacena, gubser1, witten1, review}. 
In this paper, we focus on
the 
CFT dual of the simple RS model with scalar QED 
in the bulk. 
In this case, 
the sub-leading large-$N$
corrections 
to the (external) gauge coupling
in the dual CFT are due to the loop of a fundamental scalar 
which is {\em external} to the CFT and
{\em sub-leading} part of the vacuum polarization (i.e., running)
due to CFT charged matter.
We point out
that these two effects are 
{\em comparable} in size and that the 
latter
(i.e., sub-leading part of 
CFT loop
correction) 
{\em is} 
sensitive to the strong CFT dynamics.
Thus, 
the loop
corrections to the gauge coupling in this RS model 
are difficult to compute
using
the dual theory.
 

The paper is organized as follows. In section \ref{review}, we define the 
RS model
(the dual of which is to be studied), namely scalar QED in bulk,
and give expressions
for the classical and one-loop corrected low-energy gauge coupling in this 
model. In
section \ref{RS2dual}, we 
discuss dual of the RS2 model and present the
central point mentioned above, in particular, the 
limit of induced kinetic term
for the gauge field external to the CFT.
Section \ref{RS1dual} deals with CFT dual of the RS1 model. Specifically,  
in section \ref{mixing}, we discuss
kinetic mixing between
fields external to CFT and CFT bound states
and 
also
give a dual interpretation of coupling of KK modes. 
In section 
\ref{subleading}, we discuss low-energy gauge coupling in the dual CFT
including the sub-leading large-$N$ effects (which are dual to
loop corrections on RS side). We conclude in section \ref{conclude}.

\section{Review of RS model with bulk gauge fields}
\label{review}

For simplicity, we consider (massless)
scalar QED in bulk -- extension
to non-abelian gauge fields in bulk should be straightforward. The bulk 
$5D$ action is:
\begin{equation} 
S_{bulk} = \int d^4 x \; r_c d \theta \sqrt{-G} \left(
- \frac{1}{4 g_5^2} F_{MN} F^{MN} + G^{M N} 
D_M \phi \left( D_N \phi \right)^{\dagger}
\right).
\label{Sbulk}
\end{equation}
$A_{\mu}$ and $\phi$ are taken to be orbifold-even while
$A_5$ is taken to be orbifold-odd.

In general, there are
brane-localized kinetic terms for gauge and scalar field: 
even if absent at tree-level,
these are generated 
by bulk loop effects 
due 
to breaking of translation
invariance by orbifold
\cite{georgi}. Thus, the brane action is (in addition
to action for brane-localized fields):
\begin{eqnarray} 
S_{ UV (IR) } & = & \int d^4 x \sqrt{ -g_{ UV (IR) } } 
\left (- \frac{1}{4} \tau_{UV(IR)}
F_{\mu \nu} F^{\mu \nu} + \sigma_{ UV(IR) } \left( D_{\mu} \phi
\right) ^{\dagger} 
D^{\mu} \phi \right ).
\label{Sbrane}
\end{eqnarray}
We assume that these couplings are perturbations to the bulk couplings, 
for example, they are
of same order as one-loop
processes involving bulk couplings.

The zero-mode of the gauge field
has a {\em flat} profile and hence the 
$4D$ low energy effective gauge coupling at {\em
classical} level is
\begin{eqnarray}
\frac{1}{g^2_{ 4 } } &=& \tau_{UV} + \tau_{IR} +  \frac{\pi r_c}{g^2_5} 
\nonumber \\
 & = & \tau_{UV} + \tau_{IR} + \frac{ \log \big[ O (M_4) / \hbox{TeV}
\big] }{ k g_5^2 }
\label{0modetree}
\end{eqnarray}
(using
$k \pi r_c = \log \big[ O \left( M_4 \right) / \hbox{TeV} \big]$).
From reference \cite{us}, the low energy gauge coupling
at the {\em one-loop} level (for $q \ll$ TeV) is (also, see 
\cite{pomarolprl, lisa, choi1, gr, contino, deconstruct, choi2}):
\begin{eqnarray}
\frac{1}{ g_4^2(q) } & = & \tau_{R~UV}(k) + \tau_{R~IR}(k) +  
\frac{\pi r_c}{g^2_{5~R}(k)} \nonumber \\
 & & + b_4 \left( \log \frac{k}{q} + \xi k \pi r_c + O(1) \right), 
\label{0modeloop}
\end{eqnarray}
where $\xi \sim O(1)$ and $\tau_{R~ UV(IR)}(k) 
\equiv b_4/4 \; \log \left( \Lambda/k \right) + 
\tau_{UV(IR)}$, $1/g_{5~R}^2(k) 
\equiv c \left( \Lambda - k \right) + 1/g_5^2$ are renormalized
couplings. $b_4 = 1 / \left( 24 \pi^2
\right)$ is the $4D$ $\beta$-function coefficient
of a charged scalar and $c$ is of order a $5D$ loop factor.

\section{Duality for RS2 with bulk gauge fields}
\label{RS2dual}
We now discuss the $4D$ CFT dual of this model.
The (usual) AdS/CFT correspondence 
\cite{maldacena, gubser1, witten1, review} suggests that any $5D$ gravity
theory on 
(infinite) 
AdS$_5$ is dual to some $4D$ CFT. In particular, 
for every $5D$ bulk field, $\phi$, there corresponds  
an operator, ${\cal O}$ in the CFT and the  
value of the (bulk) field at the $4D$ boundary of
AdS$_5$ (at $z = 0$), $\phi _0$ 
acts as a source for the operator. The AdS/CFT correspondence tells us that 
\begin{eqnarray} 
\int {\cal D} \psi_{CFT} \; \exp \Big[ - \left( S_{CFT} 
\big[ \psi_{CFT} \big] + \int d^4 x 
\phi_0 {\cal O} \right) \Big] 
& = & 
\int _{ \phi (z = 0) 
= \phi_0} {\cal D} \phi \; \exp \left( 
- S_{bulk} \left[ \phi \right] \right)
 \nonumber \\
 & \equiv & \exp \left( - \Gamma \left[ \phi_0 \right] \right),
\label{adscft} 
\end{eqnarray} 
%
where $S_{CFT} \big[ \psi_{CFT} \big]$ 
is the pure CFT action for CFT fields $\psi_{CFT}$
and 
$S_{bulk}$ is the bulk $5D$
action for the field $\phi$ \cite{gubser1, witten1}. The RHS is just the
effective action,
$\Gamma \left[ \phi_0 \right]$ 
of the 
(unique) 
solution for $\phi$ in bulk with $\phi$ at 
the boundary $= \phi_0$. 

Reference
\cite{susskind}
suggested that
gravity in a {\em finite} region of AdS$_5$ is dual to a CFT with a UV cut-off
related to the location of the boundary.
Since RS2 model is AdS$_5$ space cut-off by  
a Planck brane 
(i.e., with the boundary) 
at $z_{UV} > 0$,
the 
modification of the 
AdS/CFT correspondence 
in the case of 
RS2
is that 
the   
presence of the Planck brane 
corresponds
to putting a UV cut-off on the CFT of $
\sim \left( 1/z_{UV} \right)  
$ (but in 
such
a way 
that the theory remains conformal in the IR)
\cite{adscftrs2, nima, rattazzi, perez}. To be more precise, 
there exists a (unknown) reguralization scheme for the CFT 
such that 
(here, $\phi$ 
collectively
denotes
all bulk
fields)
\begin{eqnarray} 
\int _{ \Lambda_{CFT} \sim z^{-1}_{UV} }
{\cal D} \psi_{CFT} \; \exp \Big[ - \left( S_{CFT} \big[ \psi_{CFT} \big] 
+ \int d^4 x
\phi_0 {\cal O} \right) \Big] 
& = &
\int _{ \phi (z = z_{UV}) 
= \phi_0} {\cal D} \phi \; \exp \left( 
- S_{bulk} \left[ \phi \right] \right), \nonumber \\
\label{rscft1}
\end{eqnarray}
where $\Lambda_{CFT} \sim 1/z_{UV}$ in the path integral on the LHS indicates 
indicates reguralization of CFT with  
the associated mass scale being $\sim \left( 1/z_{UV} \right)$
(see, for example, reference \cite{perez}).

Since 
bulk fields evaluated on
AdS$_5$ boundary 
are dual to sources in the CFT,
the (Planck) brane-localized 
action for bulk
fields, $S_{UV}$ 
(including kinetic terms) should correspond to 
a 
{\em bare} 
action (in particular, {\em bare} kinetic terms)
for sources in the dual
CFT 
-- we will check this ansatz
in what follows. 
We 
will also show in what follows that
even if 
these bare/brane-localized kinetic terms are small (or absent),
kinetic terms for the sources will be 
``induced'' by CFT loop effects due
to the presence of a UV cut-off. This 
is dual to the fact that the boundary fields
become dynamical in RS2, {\em unlike} in the case of the boundary at
$z = 0$. 
Since
sources/boundary values of bulk fields are dynamical, we have to include 
both sides of Eq. (\ref{rscft1}) 
in a path integral over $\phi_0$ (for scalar fields, these issues
were studied in reference \cite{perez}). However, on the CFT side, we 
will still refer to these (dynamical) fields 
as ``sources'' to avoid confusion with 
CFT fields (or with the bulk fields on the RS side).
Explicitly, we have
\begin{eqnarray} 
\int {\cal D} \phi_0 \; 
\exp \left( - S_{UV} [ \phi_0 ] \right) \int _{ \Lambda_{CFT} \sim z^{-1}_{UV} }
{\cal D} \psi_{CFT} \; \exp \Big[ - \left( S_{CFT} 
\big[ \psi_{CFT} \big] + \int d^4 x
\phi_0 {\cal O} \right) \Big] 
= \nonumber \\
\int {\cal D} \phi_0 \; \exp \left( - S_{UV} [ \phi_0 ] \right)
\int _{ \phi (z = z_{UV}) 
= \phi_0} {\cal D} \phi \; \exp \left( 
- S_{bulk} \left[ \phi \right] \right)
\label{rscft2}
\end{eqnarray}
which follows from Eq. (\ref{rscft1}).
Here $S_{UV} [\phi_0 ]$ on LHS is interpreted as {\em bare} action for 
(dynamical) sources.

In this paper, we discuss aspects of the correspondence summarized in
Eq. (\ref{rscft2}) for bulk scalar QED. 
The $5D$ theory with bulk gauge fields is dual to
a $4D$ CFT with 
(unbroken) global symmetries 
\cite{witten1}. 
Then, 
the source (gauge field) couples
to the (conserved) current corresponding to a 
subgroup of this 
global 
symmetry, i.e., 
${\cal O} \sim J^{ 
global } 
_{\mu}$ and $\phi_0 (\hbox{source}) 
\sim A_{\mu} (x, z =  
\hbox{boundary})$.
In the case of
a massless bulk scalar, the corresponding
${\cal O}$ has conformal dimension $4$ 
(marginal operator) \cite{witten1}.

From the discussion above, it is clear that
kinetic terms 
for gauge and scalar field
in Eq. (\ref{Sbrane}) 
should correspond to 
a 
{\em bare} 
kinetic terms 
for sources in the dual
CFT. 
Thus, the 
CFT dual of scalar QED in the RS bulk has the following
action 
{\em at the cut-off} 
$\sim O \left( M_4 \right)$ (i.e., 
{\em bare} action) 
%
\begin{eqnarray}
S & \sim S_{CFT} + \int d^4 x \left( A_{\mu} J^{\mu} - \frac{1}{4} \tau_{ UV } 
F_{\mu \nu} F^{\mu \nu} +
\sigma_{ UV } | D \phi |^2 + \frac{1}{k} \phi {\cal O} \right). 
\end{eqnarray}
Here, 
$A_{\mu}$ is the (source) gauge field 
(``photon''), $\phi$ is the (source) charged scalar\footnote{Again, 
these source fields are dual to boundary values of bulk
AdS fields, but for notational ease, we drop the subscript $0$.}, 
${\cal O}$ 
is a dimension four CFT operator with zero anomalous dimension and  
$J^{\mu}$ is a $U(1)$ global symmetry current of the CFT. 

The dual CFT is unknown and strongly coupled.
However, one {\em can} still 
compute {\em form} of certain quantities (for example,
dependence on momenta) and also 
estimate some 
quantities in terms of $N$ (if the CFT is a large-$N$ $SU(N)$ gauge theory). 
We will check that these CFT estimates agree with
the corresponding calculations on the RS side
(which are at weak coupling and hence under control).

To begin with, 
we
include effects of
CFT loops and the $\phi$ loop 
on the $A_{\mu}$ propagator assuming that, at the cut-off,
the external gauge field is weakly coupled to the CFT, i.e.,
$1/\tau_{UV} 
\ll
16 \pi^2$.
We 
obtain, 
at energy scale $q \ll k$,  
the following expression for 
the 
$1$PI two-point function
(inverse propagator) for $A_{\mu}$ --
an 
explanation is given below
(also see discussion in references \cite{nima, gr}):
\begin{eqnarray}
\left( \eta _{\mu \nu} q^2 - q_{\mu} q_{\nu} \right) 
\left( 
\tilde{\tau}_{UV} + \big[
b_{CFT} + b_{scalar} \big] 
\Big[
\log \frac{ k
}{q} 
+ O(1) 
\Big]
\right).
\label{CFTlargeq}
\end{eqnarray}
%
The tensor
structure is fixed by current conservation. 
We assume that the
CFT is a (strongly coupled) large-$N$ $SU(N)$ gauge theory, but 
the global symmetry group
(the $U(1)$ subgroup of which is gauged) 
acts on  
fundamentals (and not, for example, adjoints) of the $SU(N)$ gauge
theory. We also assume that the number of these fundamentals is fixed
in the large-$N$ limit.
Since ``color'' of CFT acts as ``flavor'' for photon, 
running due to charged CFT fields is given by 
$b_{CFT} \log \left( \Lambda_{CFT} / q
\right)$ 
with $b_{CFT}
\sim N / \left( 16 \pi^2 \right)$ (this is the effect of the
$\langle J_{\mu} J_{\nu} \rangle$ correlator). 
This running is expressed in Eq. (\ref{CFTlargeq}) as 
$b_{CFT} \big[ \log \left( k / q \right) + O(1) \big]$ using 
$\Lambda_{CFT} \; \hbox{(UV cut-off
of CFT)} \; \sim k 
\left( \sim O \big[ M_4 \big] \right)
$. 
Here and henceforth, 
$O(1)$ refers to terms which are bounded as $q/k \rightarrow 0$. 
Although $b_{CFT}$ 
looks like
a one-loop $\beta$-function coefficient, it is {\em not}
(entirely) fixed by (gauge group) representations (i.e., quantum numbers)
of 
the CFT charged matter 
(unlike in 
one-loop
perturbation theory), 
but it 
{\em depends on the strong CFT dynamics}. However, 
the above $q$ dependence of this running
(i.e., the fact that $b_{CFT}$ is a constant)
is fixed 
by
conformal invariance.

Next, we argue that the 
running due to scalar loop 
has the form of the $b_{scalar}$ term in Eq. (\ref{CFTlargeq})
even though this includes 
{\em dressing of 
scalar propagator 
by the
$\langle {\cal O} {\cal O} \rangle
$ 
correlator
and scalar-photon vertex by the 
$\langle J_{\mu} {\cal O} {\cal O} 
\rangle
$ correlator}
(since $\phi$ couples to CFT fields through 
$\phi {\cal O}$).\footnote{We
thank Walter Goldberger 
for discussions on this point.}
Note 
that the $\langle {\cal O} {\cal O} \rangle
$ 
correlator has a quadratic divergence which cancels the $1/k$ suppression 
in the
$\phi {\cal O}$ coupling: $\langle {\cal O} (p) {\cal O} (0) \rangle
\sim N / \left( 16 \pi^2 \right) 
p^2 \big[ \Lambda_{CFT}^2 + p^2 \log \left( \Lambda_{CFT} / p
\right) + \; \hbox{finite} \big]$, where
$\Lambda_{CFT} \sim k$ (see reference \cite{perez}). 
The quadratically divergent part of this correlator
results in wavefunction renormalization
for the scalar and, in particular,
``induces'' a kinetic term for $\phi$ in the limit
$\sigma_{UV} \rightarrow 0$. So, the $\phi$ propagator
is 
\begin{eqnarray}
p^2 \left( \sigma_{UV} + \frac{N}{ 16 \pi^2 } 
\Big[ a + b \frac{p^2}{k^2} 
\left( 
\log \frac{k}{p} 
+ O(1) \right) 
\Big] \right),
\label{scalarprop}
\end{eqnarray}
where $a$ and $b$ are $O(1)$ 
constants.
A similar argument applies to the 
dressing of the scalar-photon vertex
(i.e., the $\langle J_{\mu} {\cal O} {\cal O} 
\rangle
$ correlator).
The leading order (i.e., 
quadratic divergence of) dressing of scalar propagator
is related to that of 
the scalar-photon vertex 
by the Ward
identity (i.e., gauge invariance).
The logarithmic divergence and finite part (suppressed by $\sim p^2 / k^2$) 
of CFT dressing of scalar propagator
and scalar-photon vertex can be neglected for momenta smaller than $k$.
Using these arguments, one can show that (for $q \ll k$)
running due to scalar 
has the form shown above with $b_{scalar}$ given
by $b_4$, the (one-loop) $4D$ 
beta-function \cite{gr}.

Any remaining
CFT loop corrections (at {\em sub-leading} order in a large-$N$ expansion)
are incorporated in
$\tilde{\tau}_{UV}$ (as a correction to $\tau_{UV}$).
The {\em precise} coefficients
of 
the
CFT loop corrections 
in $b_{CFT}$ and $\tilde{\tau}_{UV}$ {\em are} sensitive to the strong CFT 
dynamics.
%

Since the 
source 
is dual to the gauge/scalar field 
evaluated on
the {\em Planck brane}, 
the above $1$PI
two-point function (henceforth referred to as
``kinetic term'') for source should correspond
(on the RS side) to (inverse of) gauge/scalar  
propagator 
with both external points
on {\em Planck brane} (henceforth referred to as the
``Planck brane propagator'').

Let us see 
if 
the two sides agree.
In reference \cite{pomarolprl},
the tree-level {\em gauge} propagator in RS2 (for arbitrary external points)
was computed
using Neumann boundary condition at the Planck brane, i.e.,
{\em with no
brane-localized gauge kinetic term} ($\tau_{UV} = 0$). To compute
any gauge propagator
with $\tau_{UV} \neq 0$,
we 
have to {\em modify the (Neumann) boundary condition at the Planck brane}.
Solving the classical wave equation
of motion (including the effect of $\tau_{UV}$), we get (inverse of)
wavefunction of a continuum mode of mass $m$
{\em at Planck brane}
(for $m \ll k$) is
\begin{equation}
\left( \psi_m \right)^{-1} \sim \sqrt{\frac{m}{k}}  
\left( 
- 
\Big[
\log \left( \frac{m}{2k} \right) + \gamma 
\Big] 
+ \left( k g_5^2 \right) \tau_{UV} 
+ O \left( \frac{ m^2 }{ k^2 } \right) \right),
\label{wavefunc}
\end{equation}
%
where $\gamma \approx 0.58$ is the Euler constant. 
The Planck brane propagator (including the gauge coupling) is 
given by an integral over continuum modes
$\sim
\int ^{\sim q} dm \; 1/ q^2 \; g_5^2 \left( \psi_m \right)^2$. Here,
we neglect the contribution
from modes heavier than $\sim q$ (since these modes
give {\em contact
i.e., local} 
interactions) and neglect the mass 
$m$ in propagator for a mode lighter than $\sim q$. 
We can also solve for the propagator (Green function) {\em directly}, i.e., 
repeat the calculation of reference \cite{pomarolprl}, 
but now modifying the Neumann boundary condition
to include the effect of  
$\tau _{ UV }$.
Both methods give the following 
(inverse of) {\em tree-level} Planck brane
gauge propagator (for $q \ll k$ and ignoring the tensor structure) 
%
\begin{equation}
q^2 \left( \tau_{UV} + \frac{1}{g_5^2 k} \Big[
\log \left( \frac{2k}{q} \right) - \gamma + i \frac{ \pi }{2}  
\Big] \right) \Big[ 1 + 
O \left( \frac{ q^2 }{ k^2 } \right) \Big].
\label{Plprop}
\end{equation}
The AdS/CFT correspondence tells us that 
the 
classical AdS action is captured
by leading large-$N$ effects of the dual CFT 
\cite{maldacena, gubser1, witten1, review},
where
we can neglect $b_{scalar}$ and set $\tilde{\tau}_{UV} = \tau_{UV}$
in Eq. (\ref{CFTlargeq}). 
Then, we see that
the two sides (Eqs. (\ref{CFTlargeq}) and (\ref{Plprop})) agree {\em if}
$b_{CFT} |_{large-N} = 1/ (k g_5^2)$: 
this is consistent with
the 
fact that
$b_{CFT} |_{large-N}$ has an IR-free
sign (even if $A_{\mu}$
is non-abelian as long as rank of the gauge group is fixed in the large-$N$ 
limit) since it is dominated
by the running due to charged CFT 
matter 
which comes in complete
large-$N$ representations. 
Also, $O(1)$ term in Eq. (\ref{CFTlargeq}) has to be 
$( \log 2 - \gamma + i \pi / 2 )$ for this agreement.

Similarly, one can
compute the {\em scalar} Planck brane propagator {\em including the effect
of brane-localized kinetic term} and show that it agrees (in form) with the
scalar (source) propagator is the dual CFT, Eq. 
(\ref{scalarprop}). 

{\large \bf ``Induced'' gauge theory}:
Let us study the limit $\tau_{UV} \rightarrow 0$ on both the RS2 and 
the dual CFT sides.
On the RS2 side, 
this is 
the (simple) limit of
{\em only} bulk gauge kinetic term studied in reference \cite{pomarolprl}.
In this case, for $q \ll k$, the tree-level
{\em Planck brane} propagator is 
$ g_5^2 k / \left( q^2 \big[ 
\log \left( 2 k / q \right) - \gamma \big] \right)$ 
\cite{pomarolprl}. We see that the 
{\em classical} gauge coupling measured on Planck brane 
(defined as $q^2 \times$ propagator,
i.e., $\sim g_5^2 k / \big[ \log
\left( 2 k/q \right) - \gamma \big]$, see also reference \cite{kaloper}) 
becomes very large (i.e., the theory becomes strongly coupled)
as $q \rightarrow 2 e^{-\gamma}\; k 
\sim k$. 
Whereas, 
with $\tau_{UV} \neq 0$,
gauge coupling measured on Planck brane as $q \rightarrow k$
can be small (see Eq. (\ref{Plprop})). Also, 
even if $\tau_{UV} = 0$
at tree-level, 
bulk loops generate a (logarithmically divergent) 
brane-localized kinetic term \cite{georgi}, thus  
requiring
a brane-localized {\em counterterm} so that it is 
unrealistic to have no tree-level $\tau_{UV}$.

In the dual CFT, $\tau_{UV} \rightarrow 0$
corresponds to no (or very small) {\em bare} kinetic term 
for source -- in other words,
the limit of infinite {\em bare} gauge coupling -- so that perturbation theory
is not really valid {\em for} $q \sim k$ (this is dual to
the {\em classical} strong 
coupling behavior for $q \rightarrow k$ on RS2 side mentioned above).  
However, the limit $\tau_{UV} \rightarrow 0$ 
is a smooth
one (at the classical level) on the RS2 
side ({\em for} $q \ll k$) and so we might expect that 
the expression for the source propagator in the dual CFT
({\em for} $q \ll k$) 
also
has a smooth $\tau_{UV} \rightarrow 0$
limit.
Therefore, 
we assume that
Eq. (\ref{CFTlargeq}) 
continues to hold 
even for $\tau _{UV } 
\rightarrow 
0$. 
Then, we see that
even though 
the gauge coupling (in the dual CFT) is strong at the cut-off, it 
does become weak for $q \ll k$ 
(so that perturbation theory
is valid). 
Also, 
{\em with} $\tau_{UV} 
\rightarrow
0$ and
for $q \ll k $, 
it is obvious that 
the gauge coupling in the dual CFT
(in the large-$N$ limit and 
for 
$b_{CFT} |_{large-N} = 1/ (k g_5^2)$)
agrees
with the {\em classical} 
gauge coupling measured on Planck 
brane in RS2. 

It is clear that 
{\em in the limit} $\tau_{UV} 
\rightarrow 0$,
the kinetic term
for source (or inverse gauge coupling) in the dual CFT (for $q \ll k$) is 
``induced'' by CFT loop effects.\footnote{Induced gauge theory 
has
been discussed in other contexts, see, for example,
\cite{akama}.} 
Of course, in general, $\tau_{UV} \neq 0$ so that 
{\em bare} gauge coupling (in
the dual CFT) can be weak 
(i.e.,
perturbation theory is valid {\em even
at the cut-off}) and 
source kinetic term (for $q \ll k$) is not fully induced
by CFT loop correction.
In fact, since the source kinetic term generated 
by the CFT loop is divergent, 
we  
have to add such a counterterm (i.e.,
add $\tau_{UV}$)
to cancel this divergence
(i.e., cut-off dependence). 
 
We now briefly discuss other aspects of this 
correspondence which illustrate
the fact that {\em classical} effects on AdS side are dual to
{\em quantum} (running) effects in CFT. 

For a fixed $q$ (and $\tau_{UV}$), 
if the UV cut-off of CFT, $\Lambda_{CFT} \rightarrow 
\infty$, then 
the (infinite amount of) running due to matter fields 
results in vanishing of the propagator (or gauge coupling) 
for the source (see Eq. (\ref{CFTlargeq})). Since $\Lambda_{CFT}
\sim 1 / z_{ UV }$, on the AdS side, this corresponds to
the case where  
Planck brane is at 
the AdS boundary, i.e., $z_{UV} \rightarrow 0$ (with $k$ fixed).
Then, gauge field evaluated at  
boundary (which is dual to source in CFT) is {\em not}  
dynamical (in agreement with the CFT side). The reason
is that there is no normalizable zero-mode, whereas the continuum modes 
cannot reach the Planck brane
(due to 
{\em infinite}
potential barrier 
near the Planck brane 
in  
the analog quantum mechanics problem).
In fact, we  
can 
repeat the calculation of reference \cite{pomarolprl}
for a general position of
Planck brane 
(i.e., for {\em any} $z_{ UV }$, not necessarily $z_{ UV } = 1/k$)
and show that
the boundary gauge propagator 
(with $\tau_{ UV } = 0$ and for
$q z_{ UV } \ll 1$) is $\sim - g_5^2 k / \big[ q^2 \log  
\left( q z_{ UV } \right) \big]$ (which vanishes as $z_{ UV }  
\rightarrow 0$). This agrees 
with the source propagator in the dual CFT (Eq. (\ref{CFTlargeq}))
since
$\Lambda _{CFT} \sim 1 / z_{ UV }$.
Since source/boundary value of bulk field is {\em not} dynamical in the
limit $\Lambda _{CFT}
\rightarrow \infty$/$z_{UV} \rightarrow 0$, path integral
over these fields is {\em not} 
performed in Eq. (\ref{rscft2}) -- this is the limit
of {\em infinite} AdS (i.e., Eq. (\ref{adscft})) studied in the 
usual
AdS/CFT duality \cite{maldacena, gubser1, witten1, review}. 
 
With a Planck brane at $z_{ UV } \neq 0$, but no TeV brane,
the zero-mode is still {\em not} 
normalizable since it has a {\em flat} profile
constant in $z$ 
-- {\em both} branes are required 
to get a normalizable 
zero-mode for the gauge field, unlike for graviton (or massless scalar field), 
where {\em only} the Planck brane is required.
Nevertheless, for $q \neq 0$, some of the {\em continuum} modes 
do 
reach the Planck brane
(by 
tunneling through the {\em finite} barrier in the analog
quantum mechanics problem)
so that the Planck brane
propagator does 
not vanish.
However, 
as $q \rightarrow 0$, 
the gauge coupling measured on Planck brane 
($\sim g_5^2 k / \log
\left( k/q \right)$ for $\tau_{ UV } =0$) vanishes  
since 
there is no   
zero-mode 
(and continuum modes do not contribute as $q \rightarrow 0$). 
In the CFT dual, this corresponds to 
$\Lambda_{CFT} \sim 1 / z_{UV} \neq \infty$ and hence for $q \neq 0$, the
source propagator is 
non-vanishing (see Eq. (\ref{CFTlargeq})), whereas 
propagator for source 
vanishes as $q \rightarrow 0$   
since (as before) matter fields cause 
a gauge theory to be IR free \cite{nima, kaloper}.
As mentioned earlier, 
since the source/boundary value of bulk gauge field {\em is} 
dynamical in the case
of $\Lambda _{CFT} \neq \infty$/$z_{ UV } \neq 0$ (i.e., in the case
of the RS model),
we have to include both sides of Eq. (\ref{adscft}) in a path integral
over these fields (i.e., we have to use Eq. (\ref{rscft2})).

\section{Duality for RS1 with bulk gauge fields}
\label{RS1dual}
When we have a TeV brane (RS1),
on AdS side,
there {\em is} a (normalizable) zero-mode (which is flat). The
AdS/CFT correspondence as applied to RS1 
\cite{nima, rattazzi, perez}
says that the IR brane at $z_{IR}$
corresponds to
spontaneous breaking of conformal invariance 
at $\mu_{CFT} \sim 1 / z_{IR} \sim$ TeV. 
Also, KK modes of  
graviton (gauge boson) 
(with masses quantized in units of $1/z_{ IR } \sim$ TeV) 
are dual to massive spin-2 (spin-1) bound states of CFT while 
mass{\em less} bound states of CFT map on to
fields on the TeV brane \cite{nima}.

For $q \gg$ TeV, the kinetic term for the source
in the dual CFT is still given by Eq. (\ref{CFTlargeq}) (since for 
$q \gg$ TeV, the effect of breaking of conformal
invariance is not important).
Let us compare it to 
the RS1 side. In reference \cite{lisa}, 
the {\em tree-level} gauge propagator in RS1 (for arbitrary external points)
was computed 
using Neumann boundary condition
at {\em both} Planck and TeV branes, i.e., again {\em with} $\tau _{ UV,
IR } = 0$ --
for
$q \gg$ TeV, the 
{\em Planck brane} propagator in RS1 is
{\em same} as in that in RS2 (this propagator
obviously includes
effects of KK {\em and} zero modes). 
Thus, it agrees with source propagator in the dual
CFT (in the large-$N$ limit and {\em with} $\tau_{ UV } = 0$)
for $b_{CFT} |_{large-N} = 1 / \left( g_5^2 k \right)$. 
Again, in the limit $\tau_{ UV } \rightarrow 0$ 
studied (on the RS1 side) in reference \cite{lisa},
the 
kinetic term for source in the CFT dual is 
induced by CFT loops. Thus, (in this limit) the  
{\em CFT loop contribution}
to the source 
$1$PI two-point function
{\em corresponds (on the RS1 side) to the
contribution of KK} and {\em zero modes} to the
Planck brane propagator.

As in the case of RS2, 
we can compute the tree-level Planck brane propagator 
in RS1 {\em for} $\tau _{ UV }, \tau_{IR} \neq 0$ -- 
for $m_n \gg k$, wavefunction of KK modes at Planck brane is 
as in Eq. (\ref{wavefunc})
(up to a factor of $1/\sqrt{ z_{ IR } }$ to go from continuum
to discrete
normalization).\footnote{For
$\tau_{ UV } = 0$, these
wavefunctions were computed
in \cite{pomarolplb}.} 
Approximating the sum over (propagators of) KK modes lighter than $q$ by
an integral (which is justified if $q \gg$ TeV, i.e., splitting
between KK masses) and adding the zero-mode contribution, 
we get the same propagator as in RS2 (Eq. (\ref{Plprop}))
which again agrees with source
propagator in the dual CFT (Eq. (\ref{CFTlargeq})).

A brief comment as an aside: $\tau_{ UV }$ is necessary for  
``holographic'' RG \cite{holorg} which says that
changing UV cut-off of CFT from $k$ to $k^{\prime}$ 
(i.e., Wilsonian RG) 
corresponds to
moving the position of Planck brane ($z_{UV}$) 
from $1/k$ to $1 / k^{\prime}$.  
On AdS side, to 
keep physical $4D$ gauge 
coupling (see Eq. (\ref{0modetree}))
invariant as we move the Planck brane (with the TeV brane, i.e.,
$z_{IR}$ fixed),  
we must include a boundary gauge kinetic term
and change it: 
$\tau_{UV} \rightarrow \tau_{UV} + 1 / \left( g_5^2 k \right) 
\log \left( k / k^{\prime} \right)$
(note 
that, in general, $\pi r_c \sim 
1/k \log \left( z_{ IR } / z_{ UV } 
\right)$).\footnote{We 
can 
analyze the gauge coupling as measured on
the Planck brane (instead of $4D$ gauge coupling)
with the same result.}  
On CFT side, the 
cut-off dependence 
of CFT loop 
correction  
(to the external gauge coupling) is absorbed 
by $\tau_{UV}$ (as mentioned earlier). 

\subsection{Dual interpretation of coupling of KK modes}
\label{mixing}
Actually, there
is a subtlety in the CFT loop ``calculation'': we expect
there to be  
kinetic mixing between source 
(again, this is a dynamical field, but external to
CFT)
and spin-$1$ CFT bound states \cite{nima} 
(similar to $\gamma-\rho$ mixing in QED coupled
to QCD). For example,
if we cut the Feynman diagrams which give
the leading contributions to $\langle J_{\mu} J_{\nu} \rangle
 _{CFT}$, 
we get
(only) {\em one}-``meson'' intermediate states (as in 
the case of large-$N$ QCD -- see,
for example, reference \cite{witten2}). Thus, in the large-$N$ limit,
$\langle J_{\mu} J_{\nu} \rangle
 _{CFT}$ is a sum of {\em tree}-level diagrams in which $J_{\mu}$ creates a 
``$\rho$'' meson with an amplitude $\langle 0 | J_{\mu} | \rho \rangle$ 
which then propagates and is absorbed by $J_{\nu}$.
Since $\langle J_{\mu} J_{\nu} \rangle
 _{CFT} \propto N / \left( 16 \pi^2 \right)$, 
we get $\langle 0 | J_{\mu} | \rho \rangle \propto 
\sqrt{N} / \left( 4 \pi \right)$ for the lightest $\rho$ meson with a mass
of $\sim \mu_{CFT}$
(there might be a suppression by powers of $m_{ \rho ^{\prime} } / \mu_{CFT}$
for heavier (``$\rho^{\prime}$'') mesons). Then,
``$\gamma - \rho$'' mixing can be represented by
$\sim \sqrt{N} / \left( 4 \pi \right) F_{\mu \nu} \rho^{\mu \nu}$.

In the case of small $\tau_{UV}$ (i.e., large {\em bare} gauge 
coupling), the above kinetic mixing is important since
the CFT loop correction, which
generates this mixing, also 
induces the kinetic term for source.\footnote{Conversely,
this
mixing is small in the limit $\tau_{UV} \rightarrow \infty$, i.e., 
small bare gauge
coupling as in the case of QED (with a UV cut-off at, say, $M_4$)
coupled to QCD.} 
As a result, the 
mass{\em less} spin-$1$ state (which is dual to zero-mode of gauge field
on RS1 side) 
is mixture of source
and CFT fields -- this is
expected since zero-mode of RS1 has a 
{\em flat} profile and hence can{\em not} 
correspond
to just the source (which is dual to the gauge field
evaluated on the {\em boundary}).
Also, {\em massive} spin-$1$ states (which were pure CFT bound states in the
absence of mixing)
now contain a part of the source. 
Hence,
fields 
external to CFT
which couple to the CFT only via the source (for example, the electron
in the case of QED coupled to QCD) 
have a significant
coupling to a {\em single}
massive state \cite{nima}. 
In fact, using the above $\gamma - \rho$ kinetic mixing, we can estimate
this (``$\rho - e$'') 
coupling to be 
of order
%
\begin{equation}
\frac{ \sqrt{N} }{ 4 \pi } \times
\frac{1}{ N /  
\left( 
16 \pi^2 
\right) \times \log \left( k / 
\hbox{TeV} \right) },
\label{rhoe}
\end{equation}
%
where the first factor 
is from $\gamma - \rho$ mixing (which converts $\rho$ to $\gamma$)
and 
the second factor accounts for
$\gamma$ propagating to couple to the electron.
We have used Eq. (\ref{CFTlargeq})
(evaluated at $q \sim m_{\rho}$) for the $\gamma$ (source) propagator
with $b_{CFT} \sim N / \left( 16 \pi^2 \right)$ and $\tau_{UV} = 0$.
This 
analysis shows that the source
propagator has 
poles corresponding to masses of these states -- 
Eq. (\ref{CFTlargeq})
is strictly speaking 
{\em valid only 
for Euclidean momenta}.

This is dual to
the fact that
the coupling of a gauge KK mode (which is dual to a 
massive spin-$1$ state in the CFT) to
Planck brane fields  
(which are dual to fields 
external to CFT), given by
$g_5 \psi_{m_n}$ (see Eq. (\ref{wavefunc})), is
sizable.
For the lightest KK mode, this coupling is 
$\sim g_4 / \sqrt{k \pi r_c} \sim 
0.2 \times g_4$ (using $k \pi r_c
\sim \log \left( k / \hbox{TeV} \right)$) in the case
of
$\tau _{ UV } = 0$
\cite{pomarolplb}.
In fact, this coupling {\em agrees} with our dual CFT estimate above (Eq. (\ref{rhoe}))
since $1/g_4^2 \sim k \pi r_c / \left( g_5^2 k \right)$ (see Eq. 
(\ref{0modetree}))
and
$1 / \left( k
g_5^2 \right) = b_{CFT} |_{large-N} \sim N / \left( 16 \pi^2 \right)$.
Also, since the Planck brane propagator is a sum 
over (propagators of) these KK modes, it
has poles
at the KK masses 
-- {\em the expression for RS1 Planck brane propagator}
in Eq. (\ref{Plprop})
is also 
strictly speaking {\em valid 
for Euclidean momenta}.

The case of gauge field is to be
compared to that of gravity, where
mixing between source (which is
dynamical) and spin-$2$ CFT bound states 
is very small for $q \ll k$. 
The reason is that, in the limit of very large (or infinite) 
{\em bare} gravitational constant, 
gravity (rather inverse Newton constant) is induced
by
the {\em quadratically} 
divergent part of CFT loop correction \cite{nima, rattazzi, hawking}, 
i.e., the
$\langle T_{\mu \nu} T_{\rho \sigma} \rangle
$ correlator
(which has a form similar to the $\langle {\cal O}
{\cal O}\rangle
$ correlator given earlier). Whereas, the {\em mixing}
should be due to {\em logarithmically} 
divergent contribution and hence
is suppressed by $\sim q^2 / k^2$. 
Since the mixing is small, the source 
(which is dual to graviton
field evaluated at the {\em boundary}) is mostly the
mass{\em less} spin-$2$ eigenstate and 
{\em massive} spin-$2$ states are mostly
CFT bound states (with very small admixture of source). Thus, 
fields 
external to CFT (which couple to the CFT only via
the source) 
couple {\em weakly} to a {\em single} massive spin-$2$
state. It is also clear that since the contribution of {\em massive}
states to the source propagator (and hence
to 
a scattering process involving fields external to the CFT) 
is due to mixing, it is suppressed by $\sim q^2 / k^2$ compared
to 
the contribution
of the mass{\em less} state. 

On the RS side, this 
matches with {\em localization} of
graviton zero-mode (which corresponds to mass{\em less} spin-$2$ state 
in dual CFT) 
near the {\em Planck} brane and {\em weak} coupling
of a KK graviton mode (which is dual to {\em massive} spin-$2$ state in CFT) 
with mass $\ll k$
to Planck brane fields (which are dual to
fields external to the CFT) \cite{rs}.\footnote{As in the case of 
the gauge field, 
using the 
form of the $\langle T_{\mu \nu} T_{\rho \sigma} \rangle
$ correlator,
we can estimate the 
coupling of (lightest) massive spin-$2$ state in the dual CFT
to fields external to the CFT. 
This coupling ``agrees'' with the coupling of (lightest)
graviton KK mode to Planck brane 
fields.}
Also, the contribution of KK graviton modes
to the static potential between two 
masses
on 
the Planck brane is suppressed by $\sim 1 / \left( k r \right)^2$
compared to that of the zero-mode \cite{rs}. 
Obviously, a similar analysis holds
for the case of a massless bulk scalar.

Of course, it is difficult to compute (analytically) the
mixing between source and spin-$1$ CFT bound states
since the CFT loop involves
strong dynamics
(just as
one has to resort to, for example, lattice 
techniques
to derive $\gamma-\rho$ mixing starting from QCD). Nevertheless,
one can do the following check
(which is an extension
of the analysis of reference \cite{nima} to the case of $\tau _{UV} \neq 0$
on RS1 side).

We can
obtain the ({\em 
exact} form of) CFT loop correction
(self-energy
of photon)
from the AdS side 
{\em using the correspondence} \cite{nima}: Eq. (\ref{adscft})
implies that the 
(large-$N$ limit of)
correlator $\langle J_{\mu} J_{\nu} \rangle_{CFT}\equiv
\int {\cal D} \psi_{CFT} 
J_{\mu} J_{\nu} \exp \big[ - S_{CFT} ( \psi_{CFT} ) \big]$ 
is given by dependence of 
(tree-level) 
$5D$ action 
on boundary value of gauge field
(with brane localized coupling, $\tau_{ UV } = 0$).\footnote{This
$5D$ quantity, i.e., 
$\partial ^2 \Gamma \left[ \phi _0 \right] / \partial \phi ^2_0$
is equal to (inverse of) tree-level Planck brane propagator 
(with $\tau _{ UV } = 0$).}
Then, we can 
compare 
the 
result of resumming the source propagator
(i.e., bare 
kinetic term $\tau_{UV}$ + 
{\em exact} CFT loop)\footnote{We can show that,
in the large-$N$ limit (which corresponds to classical limit on RS1 side),
the
$\langle J_{\mu} J_{\nu} \rangle$ correlator is the same as in the 
{\em pure} CFT 
case (i.e., $\langle J_{\mu} J_{\nu} \rangle _{CFT}$ with
{\em no} propagating sources) since
the virtual effects of the sources ($A_{\mu}$ and $\phi$)
propagating (on this correlator)
are {\em sub-leading} in a large-$N$ expansion. 
Also, in the large-$N$ limit, the scalar loop can be 
neglected.} (see Eq. (39) in (preprint version of)
reference \cite{nima}) 
to
the
RS1 Planck brane propagator 
{\em as modified by 
the addition
of 
the brane-localized coupling}, $\tau_{ UV }$.
This resummed source propagator
has poles corresponding
to masses $m_n$ of RS1 KK modes 
\cite{nima}.\footnote{The 
masses of RS1 KK modes which are much lighter than $k$ 
are
the {\em same} as in the case $\tau_{ UV } = 0$.}
Away from poles (for example, with Euclidean momenta), for
$q \gg$ TeV, this source propagator (with the {\em exact} CFT loop
correction)
agrees with the
RS1 Planck brane propagator which we computed above in Eq. (\ref{Plprop})
{\em with a modification of the
(Neumann) boundary condition at the Planck brane to include the effect
of} $\tau_{UV}$.
We already mentioned
this agreement based on the {\em general} 
form of CFT loop correction (i.e., based on Eq. (\ref{CFTlargeq}) which
{\em is} reasonably 
correct away from poles).
The coupling
of a
massive spin-$2$
state 
to fields 
external to the CFT 
can be computed from the residue of the source propagator
at the pole $m_n$ (see
Eq. (44) in (preprint version of) 
reference \cite{nima}). 
One can show that it agrees with wavefunction of the RS1 KK mode
of mass $m_n$ at the Planck brane, i.e., $g_5 \psi_{m_n}$ 
which we 
computed above in Eq. (\ref{wavefunc})
(again, {\em with a
modified
boundary condition at the Planck brane to include the effect
of} $\tau_{UV}$).


What is the dual interpretation
of coupling of KK modes of gauge field to {\em TeV} brane fields?
On RS1 side,
KK modes of gauge field couple strongly 
to TeV brane fields \cite{hewett, pomarolplb}: the coupling is $\sim g_4
\sqrt{k \pi r_c} \sim O(10) \times g_4$ (since $k \pi r_c \sim \log \big[
O \left( M_4 \right) /
\hbox{TeV} \big]$) for $\tau_{UV, IR} = 0$. This is dual to the 
coupling of {\em massive} CFT bound states to mass{\em less}
{\em bound} states 
(``$\rho-\pi-\pi$'' couplings) which is $\sim 4 \pi / \sqrt{N}$ in 
the large-$N$ limit
(so that $\pi-\pi$
loop contribution to $\rho$ propagator is $\sim 
1/N
$, i.e., sub-leading order in a large-$N$
expansion: see, for example, reference \cite{witten2}).
For $1 \left( k
g_5^2 \right) = b_{CFT} |_{large-N} \sim N / \left( 16 \pi^2 \right)$, 
this agrees with the coupling on RS1 side
(using the fact that $g_4 \sim 
\sqrt{g_5^2 k} / \sqrt{k \pi r_c}$ -- see Eq. (\ref{0modetree})).

\subsection{Dual interpretation of 
(one-loop corrected) low energy gauge coupling}
\label{subleading}
Finally, we discuss the low energy gauge coupling. For
$q \ll \mu_{CFT} \sim$TeV, on CFT side, we get
the following (inverse) source propagator:
\begin{eqnarray}
\left( \eta_{\mu \nu} q^2 - q_{\mu} q_{\nu} \right) \left(
\tilde{\tau}_{ UV } + 
\big[ b_{scalar} +
b_{CFT} \big] \log \frac{ O \left( M_4 \right) }{\hbox{TeV}} + 
\tilde{\tau}_{ IR } +
b_4 \log \frac{ \hbox{TeV} }{q} \right)
\label{0modecft}
\end{eqnarray}
since
running 
(and dressing) due to CFT fields stops at $\mu_{CFT} \sim$ TeV,
where the conformal invariance is broken 
--
at this scale, there can be threshold effects 
$\sim \tilde{\tau}_{ IR }$.
These threshold effects corresponds to 
{\em TeV} brane-localized coupling on AdS side ($\tau_{IR}$
of Eq. (\ref{Sbrane})) and hence this
notation in CFT dual: 
$\tilde{\tau}_{IR} = \tau_{IR}$ up to 
CFT 
corrections at 
{\em sub-leading} order in a large-$N$ expansion.
Below $\sim$TeV,
we have only zero-modes of photon and scalar (from the
above discussion these are mixtures of 
$\phi$, $A_{\mu}$ 
and CFT fields) so that we get the usual $4D$ running, $b_4 \; \log 
\left(\hbox{TeV}/q \right)$.

On AdS side, since mass of lightest KK state is $\sim$TeV,  
only zero-mode contributes to any gauge propagator for 
$q \ll$ TeV,  
in particular, to the propagator on the Planck brane. So,  
kinetic term for source in the CFT 
(which is dual to the propagator
on Planck brane) 
should  
match (for $q \ll$ TeV) 
with {\em zero}-mode gauge coupling of RS1 ({\em including} 
loop corrections)\footnote{This point
was also made in \cite{gr}.} and it does as follows.\footnote{This 
matching
goes through for general (i.e., not just TeV) 
values of $1/z_{IR}$ and $\mu_{CFT}$ as long 
as $1/z_{IR} \sim \mu_{CFT}$.} 
In
large-$N$ 
limit, $\tilde{\tau}_{ UV (IR) } =
\tau_{ UV (IR) }$ and we can neglect
$b_{scalar}$, $b_4$. Then,
the CFT gauge coupling in Eq. (\ref{0modecft}) 
agrees with {\em tree}-level zero-mode coupling on RS1 side 
(Eq. (\ref{0modetree})) 
again 
for 
$b_{CFT} |_{large-N} = 1/ (k g_5^2)$ \cite{nima}.
We see that
the {\em classical} $4D$ coupling
$g_4^2$ 
becoming zero
as $r_c \rightarrow \infty$ 
(see Eq. (\ref{0modetree})) 
is dual to $\mu_{CFT} \rightarrow 0$
and hence (infinite) running due to 
CFT matter fields causing gauge coupling 
to become zero
in the IR
(see Eq. (\ref{0modecft}) with $\mu_{CFT}$ 
replacing TeV) \cite{nima, kaloper} --
this is a {\em quantum} effect.

As per the
AdS/CFT correspondence,
{\em loop} effects on AdS side are dual to 
{\em sub-leading} effects in a large-$N$ expansion
in the CFT
\cite{maldacena, gubser1, witten1, review}. In this case,
the sub-leading corrections to
the
gauge coupling in the dual CFT are
the terms with $b_{scalar}$, $b_4$
and {\em sub-leading} (i.e., $O(1)/ \left( 16 \pi^2 \right)$) 
part of $b_{CFT}$ in Eq. (\ref{0modecft})  
and also the corrections to $\tau$'s. 
The
$b_{scalar}$ 
term 
is 
calculable 
($b_{scalar}$ is 
$b_4$
as 
mentioned earlier) and so is
$b_4 \log \left( \hbox{TeV}/q 
\right)$ 
which is the running due to zero-mode below TeV and
which matches with the same term in Eq. (\ref{0modeloop}). But, the 
{\em precise} coefficients in
the other
sub-leading CFT effects
{\em are}
sensitive to the strong CFT dynamics (and hence difficult to compute),
in particular, the {\em sub-leading} part of
$b_{CFT}$ which is clearly comparable to the $b_{scalar}$ term
(see Eq. (\ref{0modecft})).
However, using $k \pi r_c \sim \log 
\big[ O \left( M_4 \right) / \hbox{TeV} \big]$,
we see that the {\em general} 
form (i.e., up to uncalculable $O(1)$ coefficients)
of the sub-leading CFT effects  
agrees with RS1 loop
correction to the gauge coupling (see Eq. (\ref{0modeloop})) \cite{us}.
So, the dual CFT is not useful to compute
loop corrections, but serves as
a consistency check -- for example, if loop
correction on the RS1 side is some other
power of $k \pi r_c$, then there is no dual CFT interpretation of RS1
loop effects.

\section{Conclusions}
\label{conclude}

The RS model with 
gauge fields in the bulk is dual, in the
sense of the AdS/CFT correspondence, to a $4D$ CFT with
global symmetries of CFT gauged by
an {\em external} gauge field.
On the RS side, it is 
technically natural to have small brane-localized kinetic terms (this is the
limit often studied
in the literature). In this paper,
we have shown that this limit is dual to no/small {\em bare} kinetic term for
gauge field coupled to the CFT global current and
hence the kinetic term for this gauge field is 
{\em induced} by CFT loops.

The induced kinetic term results in kinetic mixing between external gauge field and
CFT bound states (similar to $\gamma - \rho$ mixing in QED
coupled to QCD). We showed how this understanding
facilitates a dual
interpretation of {\em flat}-profile of gauge field zero-mode (in contrast
to {\em localization} of graviton zero-mode) and also of couplings of KK modes
to brane-localized fields. 
If the standard model (SM) 
gauge fields are in the RS bulk, then there are stringent lower limits
on masses of gauge KK modes due to their contributions to
compositeness and precision electroweak observables \cite{hewett, pomarolplb, others}.
In the future, we will use the dual interpretation of couplings of KK modes
developed in this paper to obtain 
{\em dual} interpretations of these
phenomenological constraints. 

We also pointed out that loop corrections in RS
are dual to {\em sub}-leading effects in a large-$N$ expansion in the dual CFT
which consist of two {\em comparable} contributions: (a) loops
of fields {\em external} to the CFT and (b) the {\em sub}-leading 
large-$N$ part of {\em pure} CFT loops -- the latter {\em is} sensitive to the
strong CFT dynamics. 
In this paper,
we studied the CFT dual of 
scalar QED in the bulk of RS, where this (external) scalar loop is calculable
{\em in spite} of dressing due to CFT interactions.
It might be interesting to study 
the CFT dual of {\em non}-abelian gauge theory in the bulk of RS -- for example, to see  
if the analogous
(external) gauge field loop is calculable.
 
As this paper 
was in preparation, reference \cite{contino} appeared which has some 
overlap
with our discussion of loop corrections in RS being dual to
sub-leading CFT effects.

\section*{Acknowledgments}

K.~A.~ is supported by the Leon Madansky Postdoctoral
fellowship and by
NSF Grant \\
P420D3620414350. A.~D.~ is supported by NSF Grants P420D3620414350 
and
P420D3620434350. 
We thank Raman Sundrum for many useful discussions and 
for
reading the manuscript.
We also thank Enrique Alvarez,
Csaba Csaki, Walter Goldberger, Gregory Gabadadze and 
Alex Pomarol for discussions, 
Csaba Csaki for encouraging us to write up this work and 
the Aspen Center
for Physics for hospitality during part of this work.


\begin{thebibliography}{99}

\bibitem{rs}L.~Randall and R.~Sundrum, hep-ph/9905221, 
Phys. Rev. Lett. 83, 3370 (1999) and
hep-th/9906064, Phys. Rev. Lett. 83, 4690 (1999).   

\bibitem{maldacena}J.~Maldacena, hep-th/9711200, 
Adv. Theor. Math. Phys. 2, 231 (1998).

\bibitem{gubser1}S.~Gubser, I.~Klebanov and A.~Polyakov, 
hep-th/9802109, Phys. Lett. B 
428, 105 (1998).

\bibitem{witten1}E.~Witten, hep-th/9802150, Adv. Theor. Math. Phys. 2, 
253 (1998).

\bibitem{review}For a review of AdS/CFT
correspondence, see O.~Aharony, S.~Gubser, J.~Maldacena,
H.~Ooguri and Y.~Oz, hep-th/9905111, Phys. Rept. 323, 183 (2000). 

\bibitem{adscftrs2} 
J.~Maldacena, unpublished remarks;
H.~Verlinde, hep-th/9906182, Nucl. Phys. B 580, 264 (2000);
E.~Witten, ITP Santa Barbara conference
`New Dimensions in Field Theory and String Theory',
{\tt http://www.itp.ucsb.edu/online/susy\_c99/discussion};
S.~Gubser, hep-th/9912001, Phys. Rev. D 63, 084017 (2001).

\bibitem{nima}N.~Arkani-Hamed, M.~Porrati and L.~Randall,  
hep-th/0012148, JHEP 0108, 017 (2001). 

\bibitem{rattazzi}   
R.~Rattazzi and A.~Zaffaroni, hep-th/0012248,
JHEP 0104, 021 (2001).

\bibitem{perez}M.~Perez-Victoria, 
hep-th/0105048, JHEP 0105, 064 (2001).

\bibitem{hewett}H.~Davoudiasl, J.~L.~Hewett and T.~G.~Rizzo,
hep-ph/9911262, Phys. Lett. B 473, 43 (2000).

\bibitem{pomarolplb}A.~Pomarol, hep-ph/9911294, Phys. Lett. B 486, 153 (2000).

\bibitem{others}For
more 
phenomenology of gauge fields in the bulk of RS1,
see S.~Chang, J.~Hisano, H.~Nakano, N.~Okada and M.~Yamaguchi,
hep-ph/9912498, Phys. Rev. D 62, 084025 (2000); T.~Gherghetta
and A.~Pomarol, hep-ph/0003129, Nucl. Phys. B 586, 141 (2000);
S.~J.~Huber and Q.~Shafi,
hep-ph/0005286, Phys. Rev. D 63, 045010 (2001); 
H.~Davoudiasl, J.~L.~Hewett and T.~G.~Rizzo, hep-ph/0006041, 
Phys. Rev. D 63, 075004 (2001)
and hep-ph/0006097, Phys. Lett. B 493, 135 (2000); 
S.~J.~Huber, C-A.~Lee and Q.~Shafi,
hep-ph/0111465, Phys. Lett. B 531, 112 (2002);
J.L. Hewett, F.J. Petriello and T.G. Rizzo, hep-ph/0203091, 
JHEP 0209, 030 (2002); 
C.~Csaki, J.~Erlich and J.~Terning, hep-ph/0203034; G.~Burdman, hep-ph/0205329.
For gauge coupling renormalization
with gauge fields in bulk of RS1, see
\cite{pomarolprl, lisa, choi1, gr, us, contino, deconstruct, choi2}.

\bibitem{hawking}S.~Hawking, J.~Maldacena and A.~Strominger, 
hep-th/0002145, JHEP 0105, 
001 (2001).

\bibitem{georgi}H.~Georgi, A.~Grant and G.~Hailu, hep-ph/0012379,  
Phys. Lett. B 506, 207 (2001).

\bibitem{us}K.~Agashe, A.~Delgado and R.~Sundrum, hep-ph/0206099, to 
be published in Nucl. Phys. B.

\bibitem{pomarolprl}A.~Pomarol, 
hep-ph/0005293, Phys. Rev. Lett. 85, 4004 (2000).  

\bibitem{lisa}L.~Randall and M.~Schwartz,  
hep-th/0108114, JHEP 0111, 003 (2001) and 
hep-th/0108115, Phys. Rev. Lett. 88, 081801 (2002).

\bibitem{choi1}Gauge coupling
renormalization in {\em supersymmetric} warped compactifications is 
discussed in
K.~Choi, H.~D.~Kim and Y.~W.~Kim, hep-ph/0202257;
K.~Choi, H.~D.~Kim and I-W.~Kim, hep-ph/0207013.

\bibitem{gr}W.~Goldberger and I.~Rothstein, hep-th/0204160 and hep-th/0208060.

\bibitem{contino}R. Contino, P. Creminelli and E. Trincherini, hep-th/0208002.

\bibitem{deconstruct}Gauge coupling
renormalization in ``deconstructed'' RS1 is discussed in
A.~Falkowski and H.~D.~Kim, hep-ph/0208058, JHEP 0208, 052 (2002); 
L.~Randall, Y.~Shadmi and N.~Weiner,
hep-th/0208120.

\bibitem{choi2}K.~Choi and I-W.~Kim, hep-th/0208071.

\bibitem{susskind}L.~Susskind and E.~Witten, hep-th/9805114.

\bibitem{kaloper}N.~Kaloper, L.~Susskind and E.~Silverstein, hep-th/0006192,
JHEP 0105, 031 (2001).

\bibitem{akama}K.~Akama and T.~Hattori, hep-ph/9607331, 
Phys. Lett. B 392, 383 (1997);
G.~Veneziano, hep-th/0110129, JHEP 0206, 051 (2002).

\bibitem{holorg}M.~Porrati and A.~Starinets, 
hep-th/9903085, Phys. Lett. B 454, 77 (1999); 
V.~Balasubramanian and P.~Kraus, hep-th/9903190, Phys. Rev. Lett.
83, 3605 (1999); J.~de Boer, E.~Verlinde and H.~Verlinde,
hep-th/9912012, JHEP 0008, 003 (2000); E.~Verlinde and H.~Verlinde, 
hep-th/9912018, JHEP 0005, 034 (2000).

\bibitem{witten2}E.~Witten, Nucl. Phys. B 160, 57 (1979).

\end{thebibliography}
\end{document}